\documentclass[12pt]{amsart}
\usepackage{amssymb}
\numberwithin{equation}{section}
\newtheorem{thm}{Theorem}[section]

\renewcommand\d{\partial}

\newcommand\beq{\begin{eqnarray}}
\newcommand\eeq{\end{eqnarray}}

\usepackage[mathscr]{eucal} 
\newcommand{\scri}{{\mathscr I}}
\newcommand{\tM}{{\tilde M}}
\newcommand{\tg}{{\tilde g}}

\title[Asymptotically Simple Spacetimes]
{Some Global Results for\\ Asymptotically Simple Spacetimes}

\author{Gregory J. Galloway}
\thanks{Supported in part by NSF grant \#
   DMS 0104042.}
\address{University of Miami, Department of Mathematics and \newline  Computer Science,
Coral Gables, FL, 33124}
\email{galloway@math.miami.edu}

\begin{document}

\maketitle

\section{Introduction}

The aim of this paper is to present some rigidity results for asymptotically simple spacetimes.
By a \emph{spacetime} we mean a smooth connected time oriented Lorentzian manifold $(M,g)$ of dimension 
$\ge 3$, having signature $(-+\cdots+)$.  We use standard notation for causal theoretic notions, e.g.,
for $A\subset M$, $I^+(A,M)$, the timelike future of $A$ in $M$ (resp., $I^-(A,M)$, the timelike past of $A$ in
$M$), is the set of all points  that can be reached from $A$ by a future directed (resp. past directed)
timelike curve in $M$.  For other basic definitions and results in spacetime geometry and causal theory we refer the
reader to the standard references \cite{HE,P2}.  

Penrose's treatment of infinity \cite{P1,PR} in asymptotically flat spacetimes (and spacetimes 
with other asymptotic structures, as well) is based on
his notion of \emph{asymptotic simplicity}.  
A $4$-dimensional, chronological spacetime $(M,g)$ is
asymptotically simple
provided  there exists a smooth spacetime-with-boundary $(\tM,\tg)$
such that,
\begin{enumerate}
\item[(a)] $M$ is the interior of $\tM$, and hence, $\tM= M\cup\scri$, where $\scri = \partial \tM$,
\item[(b)] $\tg = \Omega^2g$, where $\Omega$ is a smooth function on $\tM$ such that (i) $\Omega > 0$
on $M$ and (ii) $\Omega = 0$ and $d\Omega \ne 0$ along $\scri$, and
\item[(c)] every inextendible null geodesic in $M$ has a past and future end-point on~$\scri$. 
\end{enumerate}

Condition (c) is a strong global assumption, which implies that $M$ is null geodesically complete, and ensures that
$\scri$ includes all of null infinity.  But it also rules out spacetimes with singularities, and black holes, etc. 
To treat such cases, condition (c) must be suitably weakened, cf. \cite{HE,Wa}.   

In this paper, we will be primarily interested in asymptotically simple spacetimes which obey the
vacuum Einstein equation with cosmological term,
\beq\label{eeqn}
{\rm Ric} = \lambda\, g\,.
\eeq
If $\lambda =0$ (the asymptotically flat case) then $\scri$ is necessarily a smooth null hypersurface,
which decomposes into two parts, $\scri^+$, future null infinity, and $\scri^-$, past null infinity. 
If $\lambda > 0$ (the asymptotically de Sitter case) then we have a similar decomposition, except
that $\scri$ is spacelike.

We present here a uniqueness theorem for Minkowski space and for de Sitter
space associated with the occurrence of \emph{null lines}.  A null line in a spacetime
$(M,g)$ is an inextendible null geodesic which is globally achronal (meaning that no two
points can be joined by a timelike curve).  Arguments involving null lines have arisen in many situations,
such as the Hawking-Penrose singularity theorems \cite{HE},  
results on topological censorship \cite{FSW,GSSW}, the Penrose-Sorkin-Woolgar approach  
to the positive mass theorem \cite{PSW,W} (and related results  on gravitational time delay
\cite{GW}), and most recently results concerning the AdS/CFT Correspondence \cite{GSW}.  

Every null geodesic in Minkowski space and de Sitter space is a null line.
At the  same time, each of these spacetimes, and indeed any spacetime satisfying equation~(\ref{eeqn})
(regardless of the sign of $\lambda$)
obeys the \emph{null energy condition}, ${\rm Ric}\,(X,X) =R_{ij}X^iX^j\ge 0$, for all null vectors $X$. 
In general, it is difficult for complete null lines to exist in spacetimes
which obey the null energy condition. The null energy condition tends to focus
congruences of null geodesics, which can lead to the occurence of null conjugate points.
A null geodesic containing a pair of conjugate points cannot be achronal.  Thus we expect
a spacetime  which satisfies the null energy condition  and which contains a complete null line to be
special in some way. The following theorem supports this point of view.

\begin{thm} Suppose $M$ is an asymptotically simple spacetime satisfying the vacuum Einstein
equation (\ref{eeqn}) with $\lambda \ge 0$. If $M$ contains a null line then $M$ is isometric to Minkowski space
(if $\lambda = 0$) or de Sitter space (if $\lambda >0$).  
\end{thm}

Let us give an interpretaton of Theorem 1 in terms of the initial value problem for the vacuum Einstein equation,
in the case $\lambda >0$.  According to the fundamental work of Friedrich \cite{F1}, the set of asymptotically
simple solutions to (\ref{eeqn}), with $\lambda >0$, is open in the set of all maximal globally hyperbolic
solutions with compact spatial sections. 
Thus, by Theorem 1, in conjunction with the
work of Friedrich, a sufficiently small perturbation of the Cauchy data on a fixed Cauchy hypersurface in de
Sitter space  will in general destroy all the null lines of de Sitter space, i.e.,
the resulting spacetime that develops from the perturbed Cauchy  data will not contain any null lines.
While one would expect many of the null lines
to be destroyed, it is somewhat surprising that \emph{none} of the null lines persist.   
One is tempted to draw a similar conclusion in the $\lambda =0$ case. However, the nonlinear
stability of asymptotic simplicity in this case has not been established, and, indeed may not hold, cf.,
\cite{F3,F4} for further discussion.
In fact at present, Minkowski space is the only  known asymptotically simple solution to (\ref{eeqn})
with $\lambda =0$ (though evidence is mounting that there are solutions distinct from, but in a suitable sense,
close to Minkowski space; by Theorem 1, such solutions would have no null lines).
Finally, we note that if $M$ is an asymptotically simple and de Sitter spacetime without any  null lines
then the sets $\d J^+(p)$ will be compact for all
$p\in M$ sufficiently close to $\scri^-$.  As all such sets in de Sitter space
are noncompact, this  further serves to illustrate the fragile
nature of the causal structure of de Sitter space (see also \cite[Corollary 1]{GW}).

Theorem 1 is a consequence of the \emph{null splitting theorem} obtained in \cite{G}.  This latter
result establishes the rigidity in general of null geodesically complete spacetimes which obey
the null energy condition.  We discuss this result in the next section. In section 3 we present
the proof of Theorem 1.  For results of related interest concerning the rigidity of asymptotically
simple spacetimes, see, for example, \cite{AG,F2,M}. 


\section{The Null Splitting Theorem} 

The statement of the null splitting theorem involves null hypersurfaces.  We
recall briefly some facts about the geometry of null hypersurfaces; see e.g., \cite{G,HE,P2} for more
detailed discussions from slightly varying points of view.

A smooth null hypersurface in a spacetime $(M,g)$ is a smooth co-dimension one submanifold $S$ along which
the Lorentz metric g is degenerate.  Hence, $S$ admits a smooth future directed null vector field $K$, which
is unique up to a positive scale factor.  As is well-known, the integral curves of $K$, when suitably 
parameterized, are null geodesics, and are referred to as the null generators of~$S$. 

The \emph{null expansion tensor} (or \emph{null second fundamental}) $\Theta$ of $S$ measures the 
variations in the spatial separation of the null generators of $S$.  Let $TS/K$ denote the tangent 
vectors to $S$ mod $K$; for each $p\in S$, $T_pS/K = \{\bar X: X\in T_pS\}$, where 
$\bar X= \{Y\in T_pS: Y=X \mbox{ mod } K\}$.
$TS/K$ is an $n-2$ dimensional vector bundle over $S$ ($n={\rm dim}\, M$). For $\bar X$, $\bar Y\in T_pS/K$,
define $h(\bar X,\bar Y) =g(X,Y)$; $h$ is a well-defined smooth Riemannian metric on $TS/K$.  
$\Theta$ is defined as, $\Theta(\bar X, \bar Y) = g(\nabla_XK,Y)$, where $\nabla$ is the Levi-Civita
connection of $(M,g)$; $\Theta$ is a
well-defined symmetric bilinear form on $TS/K$, unique up to the scaling of $K$. By tracing $\Theta$ with respect
to $h$, we obtain the null expansion scalar $\theta = h^{ab}\theta_{ab}$, which measures the divergence of the
null generators towards the future.  Along an affinely parameterized null generator, $s\to\eta(s)$, the null
expansion
$\theta =\theta(s)$ satisfies (provided $K$ is appropriately scaled) the Raychaudhuri equation for an irrotational
null congruence,
\beq\label{ray}
\frac{d\theta}{ds}= -{\rm Ric}\,(\eta',\eta')- \sigma^2-\frac1{n-2}\theta^2\,, 
\eeq
where $\sigma^2 = \sigma_{ab}\sigma^{ab}$ , and $\sigma_{ab}$ is the shear tensor, 
$\sigma_{ab} = \theta_{ab}-\frac1{n-2}\theta$.

We say that $S$ is \emph{totally geodesic} if and only if the expansion tensor vanishes, $\theta_{ab} \equiv0$,
or, equivalently, if and only if the expansion scalar and shear vanish, $\theta \equiv 0$, $\sigma_{ab}
\equiv0$.  This has the usual geometric meaning: A geodesic in $M$ starting tangent to a totally geodesic null 
hypersurface $S$ remains in $S$. Null hyperplanes in Minkowski space are totally geodesic, as is the
event horizon in Schwarzschild spacetime. 

We now state the null splitting theorem.

\begin{thm} Let $M$ be a null geodesically complete spacetime which obeys the null energy condition,
${\rm Ric}\,(X,X) =R_{ij}X^iX^j\ge 0$, for all null vectors $X$.  If $M$ admits a null line $\eta$ then
$\eta$ is contained in a smooth achronal edgeless \emph{totally geodesic} null hypersurface $S$.  
\end{thm} 

The simplest illustration of Theorem 2 is Minkowski space: Each null line $\ell$ in Minkowski space
is contained in a unique null hyperplane $\Pi$.  

We make some comments about the proof, which is based 
on a maximum principle for $C^0$ null hypersurfaces, see \cite{G} for details.
First, by way of motivation, note that the null plane
$\Pi$ above can be realized as the limit of the future null cone $\partial I^+(x)$ as $x$ goes to past
null infinity along the null line $\ell$.   $\Pi$ can also be realized as the limit of the
past null cone  $\partial I^-(x)$ as $x$ goes to future
null infinity along the null line $\ell$.  In fact, one sees that $\Pi = \partial I^+(\ell)=
\partial I^-(\ell)$.

Thus, in the setting of Theorem 2, consider the \emph{achronal boundaries} $S_+= \partial I^+(\eta)$ 
and $S_-= \partial I^-(\eta)$.  By standard causal theoretic results \cite{P2}, $S_+$ and $S_-$ are
achronal, edgeless, $C^0$ (but in general not smooth) hypersurfaces in $M$.  Since $\eta$ is achronal, 
it follows that $S_+$ and $S_-$ both contain $\eta$. For simplicity, assume $S_+$ and $S_-$ are connected 
(otherwise restrict attention to the component of each containing $\eta$).  The proof then consists of showing
that $S_+$ and $S_-$ agree and form a smooth totally geodesic null hypersurface.  We give some indication
as to how this works.  

By further properties of achronal boundaries (see especially \cite[Lemma 3.19]{P2}),
each point $p\in S_-$ is the past end point of a null geodesic contained in $S_-$ which is future inextendible
in $M$, and hence future  complete, i.e., $S_-$ is a $C^0$ null hypersurface ruled by future complete null
geodesics.  Similarly, $S_+$ is a $C^0$ null hypersurface ruled by past complete null
geodesics.  Now suppose $S_+$ and $S_-$ are actually \emph{smooth} null hypersurfaces.
Then by standard arguments, based on Raychaudhuri's equation (eq. (\ref{ray})), $S_-$ must
have null expansion $\theta_-\ge 0$.  Indeed, if  $\theta_-<0$ at some point $p$ then the
null generator through $p$ would encounter a null focal point in a finite affine parameter time to the future,
forcing the generator to leave $S_-$, which is impossible.   (This is the basis of the proof of
the black hole area theorem.)  Time-dually, $S_+$ has null expansion $\theta_+\le 0$.
Now, let $q$ be a point on both $S_+$ and $S_-$.  By simple causal considerations
one observes that in the vicinity of $q$, $S_+$ lies to the future side of $S_-$.  Using the
inequalities $\theta_+\le 0\le \theta_-$, and the fact that the null expansion scalar can be expressed
as a second order quasi-linear elliptic operator acting on some function related to the
graph of the null hypersurface, one can apply the strong maximum principle to conclude that
$S_+$ and $S_-$ agree near $q$, c.f., \cite[Theorem II.1]{G}.  Thus, the nonempty set 
$S_-\cap S_+$ is both open and closed in $S_-$, and in $S_+$. Hence $S_- =S_+$, and this common null
hypersurface, call it $S$, must have vanishing null expansion, $\theta\equiv 0$.  Equation (2) then implies
that the shear of the generators also vanishes, and hence $S$ is totally geodesic. 

The principal difficulty in proving Theorem 2 is in showing that the arguments of the preceding paragraph
extend to the $C^0$ setting.  In fact it is possible to show that $S_+$ and $S_-$ satisfy, 
$\theta_+\le 0 \le \theta_-$, in a  certain weak sense, namely, in the sense of smooth
\emph{support} null hypersurfaces.  Then, as an application of the weak version of the strong
maximum principle obtained in \cite{AGH} (which includes a regularity assertion), it is again possible to show that
$S_+$ and $S_-$ agree and form a smooth  null hypersurface with vanishing null expansion scalar, 
cf., \cite[Theorem III.4]{G}  

\section{Proof of Theorem 1}

A proof of Theorem 1 in the case $\lambda=0$ was given in \cite{G}.  
Here we give a proof of Theorem 1, different in a number of respects, which accommodates the case $\lambda >0$.
We also correct a mistake in the proof given in~\cite{G}.
The proof made use of the erroneous assertion that $M^- =M\cup \scri^-$ is causally simple 
(i.e., that the sets of the form $J^{\pm}(K,M^-)$ are closed for all compact $K\subset M^-$). 
As pointed out in \cite{N}, without some further assumption on $\scri$, this in principle
need not hold.  In the proof presented here we circumvent the use of causal simplicity.  

The first, and main step in proving Theorem 1, is  to show that $(M,g)$ has constant curvature.
Let $\eta$ be the assumed null line in $M$, and let $S$ be the component of $\partial I^+(\eta,M)$ containing
$\eta$.  By Theorem 2, $S$ is a smooth totally geodesic null hypersurface in $M$.  
The null line $\eta$ acquires a past end point 
$p$ on $\scri^-$ and a future end point $q$ on $\scri^+$.  We examine the structure of $S$ near $p$. (A similar
structure will hold near $q$.) For this purpose it is convenient to
extend $M^- =M\cup \scri^-$ slightly beyond its boundary.
Thus, without
loss  of generality, we may assume that $M^-$ is contained in a spacetime (without boundary) $P$ such that 
$\scri^-$ separates $P$.  It follows that $\scri^-$ is a globally achronal null hypersurface in $P$.

Observe that $I^+(\eta,M) = I^+(p,P)$, from which it follows that $\d I^+(\eta,M) = \d I^+(p,P)\cap M$
(where $\d I^+(A,X)$ refers to the boundary in $X$).  Asymptotic simplicity then implies that the generators
of $\d I^+(\eta,M)$ must have past end points on $\scri^-$ at $p$. (A generator $\gamma$ of $\d I^+(p,P)$
starting in $M$ either extends to $p$, or else is past inextendible in $P$, while remaining in $\d I^+(p,P)$.  
In the latter case,  $\gamma$ meets $\scri^-$ transversely and enters $I^-(\scri^-,P)$, violating the
achronality of
$\scri^-$.)\, The generators of $S\subset
\d I^+(\eta,M)$ must coincide with those of $\d I^+(\eta,M)$, and hence have past end points on $\scri^-$
at $p$, as well.  Let $U$ be a convex normal neighborhood of $p$, and let $A$ be the ``null cone" in $U$
generated by the future directed null geodesics in $U$ emanating from $p$.  $A$ is a smooth null hypersurface
in $U$, except for the conical singularity at $p$. We can choose a smooth spacelike hypersurface in $U$, passing
slightly to the future of $p$, which meets $A$ in a $2$-sphere $\Sigma$.  
By invariance of domain, and the fact that $S$ is closed in $M$, it is easily seen that the subset    
$\Sigma \cap S$ of $\Sigma\cap M$ is both open and closed in $\Sigma\cap M$. It follows that all the null
geodesics forming $A$ correspond  precisely to the generators of $S$, except, in the asymptotically
flat case, for the generator $\gamma_p$ of $\scri^-$ with past end point $p$.  Note in particular, this implies
that $S = \d I^+(\eta,M)$.  

Now, set $N_p = S\cup \gamma_p$ in the asymptotically flat case, and $N_p = S\cup \{p\}$\, in the 
asymptotically de Sitter case.  From the above, $N_p$ is made up of all the future inextendible null geodesics in
$P$ emanating from $p$.  In the asymptotically de Sitter case, $\scri^-$ is spacelike, so 
$\d I^+(p,P)$ meets $\scri^-$ only at $p$.  It follows from statements above that $N_p=\d I^+(p,P)$. 
Suppose, in the asymptotically flat case, that $\d I^+(p,P)$ meets $\scri^-$ at a point $x$, say.  Then
there exists a  null geodesic $\gamma\subset \d I^+(p,P)$ with future end point $x$ which either
extends in the past to $p$ or is past inextendible in $M\cup \scri^-$.  In the latter case,   Lemma 4.2 
in \cite{N} implies that $M\subset  I^+(\gamma)$, from which it follows that $M \subset I^+(p,P)$. 
But this contradicts the achronality of $\eta$.  It follows that $\d I^+(p,P) \cap \scri^- = \gamma_p$,
and again we have $N_p=\d I^+(p,P)$.
Thus, $N_p$ is an achronal boundary, and hence
is an achronal edgeless $C^0$ hypersurface in $P$. Moreover, since the future
directed null geodesics in $P$ emanating from $p$
do not cross and, being achronal, do not have points conjugate to
$p$, $N_p\setminus\{p\}$ is the diffeomorphic image under the exponential map $\exp:{\mathcal O}\subset T_pP\to P$
of $(\lambda^+_p\setminus\{0\})\cap{\mathcal O}$, where $\lambda^+_p$ is the future null cone in $T_pP$ and 
${\mathcal O}$ is the maximal open set on which $\exp$ is defined.  Thus, apart from the conical 
singularity at $p$, $N_p$ is a smooth null hypersurface. 
We are justified in thinking of $N_p$ as the future null cone in $P$ at $p$. 

The shear tensor $\sigma_{ab}$ of $S$ in the physical metric $g$ vanishes, and so, since the shear is
a conformal invariant, the shear tensor $\tilde\sigma_{ab}$ of $N_p\setminus\{p\}$ vanishes
in the unphysical metric $\tilde g$.  Then by the well-known propagation equation for the shear tensor 
\cite[Eq. 4.36]{HE}, the components $\tilde C_{a0b0}$  (with respect
to an appropriately chosen frame in which $e_0$ is aligned along the generators)
of the conformal tensor of $\tilde g$ vanish on $N_p\setminus\{p\}$.  We can now invoke an argument
of Friedrich \cite{F2} used in essentially the same situation to prove a cosmic no-hair theorem.  By
use of the regular conformal field equations, specifically the divergencelessness of the
rescaled conformal tensor,
$$
\nabla_id^i_{jkl} = 0\,, \qquad d^i_{jkl}= \Omega^{-1}\tilde C^i_{jkl} \,,
$$
Friedrich  shows that the full rescaled conformal tensor $d^i_{jkl}$  vanishes on
the future domain of dependence $D^+(N_p,P)$ of $N_p$.  Hence, the conformal tensor
with respect to the physical metric vanishes on $D^+(N_p,P)\cap M$.  Together with equation (2),
this implies that $(M,g)$ has constant curvature on $D^+(N_p,P)\cap M$.  In a precisely time-dual
fashion $(M,g)$ has constant curvature on $D^-(N_q,Q)\cap M$, where $Q$ is an extension of $M\cup \scri^+$
analogous to $P$, and $N_q$ is the past null cone in $Q$ at $q$.  

To conclude that $M$ is everywhere of constant curvature we need to show that 
$M\subset D^+(N_p,P)\cup D^-(N_q,Q)$.  Since 
$S=\d I^+(\eta,M)=\d I^-(\eta,M)$, $M$ can be expressed as the disjoint union,
\beq\label{decomp}
M = I^-(S,M) \cup S \cup I^+(S,M)\,.
\eeq
This decomposition follows from \cite[Proposition 3.15]{P2}, but can be easily shown
directly as follows.  Let $x\in M\setminus S$. There exists a curve  $\sigma$ in $M$ 
(not necessarily causal) from $x$ to $y\in S$ that meets $S$ only at $y$.  Either
$\sigma \cap I^-(S,M) \ne \emptyset$ or $\sigma \cap I^+(S,M) \ne \emptyset$.  If
the latter holds and $\sigma$ is not contained in $I^+(S,M)$, then  $\sigma$ meets $\d I^+(S,M)$ at 
some point $z \notin S$.  But since $I^+(S,M) = I^+(\d I^+(\eta,M)) = I^+(\eta,M)$,  
$z \in  \d I^+(\eta,M) = S$, which is a contradiction.  Hence, $\sigma \subset I^+(S,M)$, and so 
$x \in I^+(S,M)$.  Similarly, $\sigma\cap I^-(S,M) \ne \emptyset$ leads to $x\in I^-(S,M)$, which
establishes~(\ref{decomp}). 

We now show that each term in (3) is 
a subset of $D^+(N_p,P)\cup D^-(N_q,Q)$.  Trivially,
$S\subset N_p \subset D^+(N_p,P)$.  
Consider $I^+(S,M) \subset J^+(N_p,P)\cap M$. 
We claim that  $J^+(N_p,P)\cap M\subset D^+(N_p,P)\cap M$.  If not, then  
$H^+(N_p,P)\cap M\ne\emptyset$. Choose a point $x\in H^+(N_p,P)\cap M$, and let $\nu$ be a null
generator of $H^+(N_p,P)$ with future end point $x$.  
Since $N_p$ is edgeless, $\nu$ remains in $H^+(N_p,P)$
as it is extended into the past.  By asymptotic simplicity, $\nu$ must meet
$\scri^-$.  In fact, $\nu$ must
meet $\scri^-$ transversely (even in the asymptotically flat case, since $\nu$ starts in $M$)
and then enters $I^-(\scri^-,P)$.  But this means that
$\nu$ has left $J^+(N_p,P)$, which is a contradiction.  Hence, 
$I^+(S,M)\subset D^+(N_p,P)$, and  by the time-dual argument, $I^-(S,M)\subset D^-(N_q,Q)$.

Thus, $M$ is globally of constant curvature.  By the uniqueness of  
simply connected Lorentzian  space forms, Theorem 1 will follow once we 
show that $M$ is simply connected and geodesically complete.  In the 
asymptotically flat case, it is shown in \cite{N} that $M$ is simply connected 
(in fact, is homeomorphic to $\Bbb R^4$).  Simple connectivity in the
anti-de Sitter case may be established as follows.  Each generator of $S$
has a past end point at $p$ and a future end point at $q$. By the structure
of $S$ near $p$ and $q$ established above, it follows that $S\cup\{p,q\}$
is homemorphic to the $3$-sphere.  Perturbing  $S\cup\{p,q\}$ slightly
near $p$ and $q$, we obtain a smooth achronal $3$-sphere $S_0$ in $M$.
By an argument similar to one given above, one can show that 
$H^+(S_0,P)\cap M = H^+(S_0,Q)\cap M =\emptyset$, from which it follows
that $S_0$ is a Cauchy surface for $M$.  (Alternatively, by a suitable
modification of arguments in \cite{N}, one can show that
$M$ is globally hyperbolic. Then, since $S_0$ is compact and achronal
it must be a Cauchy surface.) \, Thus $M$ has topology $\Bbb R \times S_0$,
and so is simply connected.

It remains to show that $M$ is geodesically complete. Being 
asymptotically simple, we know that $M$ is null geodesically complete,
which is essential to the proof of full completeness.  
Let $\bar M$ denote
Minkowski space in the case $\lambda =0$, and de Sitter space in the case $\lambda >0$.
By standard results, for each $x\in M$ and $\bar x \in\bar M$, there is a neighborhood
of $x$ isometric to a neighborhood of $\bar x$.  Since $M$ is simply connected, these
local isometries can be pieced together, by an analytic continuation type argument,
to produce a local isometry $\phi: M \to \bar M$, see, for example, Theorem 8.17
in \cite{ON}.  (The assumption in \cite[Th. 8.17]{ON} that $M$ is complete is
not needed to produce the local isometry; it is used only to conclude that $\phi$
is a covering map.)  

Let $x$ be any point in $M$, and let $\bar x=\phi(x)$. Then,
from the fact that $\phi$ is a local isometry
and $M$ is null geodesically complete, it follows that any (broken)
null geodesic segment in $\bar M$ starting at $\bar x$ can be lifted uniquely via $\phi$ to 
a  (broken) null geodesic segment in $M$ starting at $x$.  From this it follows that
$\phi$ is onto: Fix $x\in M$ and let $\bar x=\phi(x)$.  Let $\bar \nu$ be a broken null geodesic
in $\bar M$ from $\bar x$ to any other point $\bar y$.  Let $\nu$ be the lift of $\bar \nu$
starting at $x$, and let $y$ be its final end point;  since $\nu$ covers 
$\bar \nu$, we must have $\phi(y) = \bar y$.

To establish geodesic completeness, it is sufficient to show that
any unit speed timelike or spacelike geodesic $\gamma:[0,a)\to M$, $t\to\gamma(t)$,
continuously extends to $t=a$.  Let $\bar\gamma=
\phi\circ\gamma$; $\bar \gamma$ can be extended to a complete  geodesic in $\bar M$ which we
still refer to as $\bar\gamma$. Fixing  $u_0\in [0,a)$ sufficiently close to $a$, one easily constructs
a $C^0$ homotopy of curves $\{\bar\sigma_u\}$, $u_0\le u < a$, from $\bar\gamma(0)$ to
$\bar\gamma(a)$, where, for each $u\in [u_0,a)$, $\bar\sigma_u$ consists of the segment  $\bar
\gamma|_{[0,u]}$,  followed by a suitably chosen two-segment broken null geodesic $\bar\nu_u$ from
$\bar\gamma(u)$ to $\bar\gamma(a)$,
such that the length of $\bar\nu_u$ (in some
background Riemannian metric) goes to zero as $u\to a$.  The broken null segments $\bar\nu_u$ can be constructed
from  the two families of null geodesics foliating a totally geodesic timelike $2$-surface containing 
$\bar\gamma|_{[u_0,a]}$.  By the nature of the lifting procedure, the homotopy $\{\bar\sigma_u\}$, $u_0\le u
<a$  can be lifted to a $C^0$ homotopy $\{\sigma_u\}$, $u_0\le u <a$, where, for each $u$, $\sigma_u$
consists of the segment  $\gamma|_{[0,u]}$,  followed by a two-segment broken geodesic
$\nu_u$ covering $\bar\nu_u$.   The curves $\sigma_u$ must have a common final end point $q$, say. Since
the length of $\nu_u$ goes to zero (in the lifted Riemannian metric) as $u\to a$, it follows that
$\gamma(u) \to q$ as $u\to a$, i.e., $\gamma$ is extendible to $q$.  This completes the proof 
of Theorem 1.

\section{Concluding Remarks}

It is possible to formulate a version of Theorem 1 applicable to anti-de Sitter space.  
However, since in this case $\scri$ is timelike, the 
characteristic initial value problem encountered in the proof of Theorem 1 would become an
initial-boundary value problem, and it would be necessary to impose boundary data (namely
the vanishing of the conformal tensor) on $\scri$.  In this case one might expect
to establish the uniqueness of anti-de Sitter space without the assumption of a null line. 

We believe that one should be able to generalize Theorem 1 in certain directions.
For example, in the asymptotically
flat case, we expect that the vacuum assumption could be weakened to allow for the presence of
matter.  For instance, if we assume spacetime
$M$ satisfies the null energy condition, and is vacuum in a neighborhood of $\scri$, then
the proof of Theorem 1 implies the existence of open sets near scri which are flat. If the vacuum
region is analytic then it follows that a neighborhood of scri is flat, and so $M$ should have
vanishing mass.  By  a suitable version of the positive mass theorem, 
$M$ should be isometric to Minkowski space.
Thus, we conjecture that, in the asymptotically flat case, the vacuum assumption can
be replaced by the null energy condition and a requirement that the spacetime Ricci tensor falls
off at an appropriate rate on approach to $\scri$.  It may also be 
possible, in the asymptotically flat case,
to prove a version of Theorem 1 for \emph{weakly}
asymptotically simple spacetimes, which allows for the occurence of black holes. By imposing suitable
conditions on the domain of outer communications $I^-(\scri^+)\cap I^+(\scri^-)$ (e.g., that the DOC
be globally hyperbolic) one might expect to be able to show that the DOC is flat.  
To accomplish this, one may be able to exploit the fact that the proof of the null splitting theorem
does not actually require the full null completeness of spacetime.

We mention in closing that the null splitting theorem has recently been  used \cite{AG} to obtain
restrictions on the topology of $\scri^{\pm}$ in asymptotically simple and de Sitter spacetimes
obeying the null energy condition.

\vspace{.2in}
\noindent
\begin{large}{\bf Acknowledgement}
\end{large}

\medskip
\noindent
This paper is based on a talk given at the Workshop on the Conformal Structure of Spacetimes,
held in T\"ubingen, April 2-4, 2001.  We wish to thank the organizers J\"org Frauendiener
and Helmut Friedrich for their gracious hospitality and support.  We also wish to thank
Piotr Chru\'sciel for comments on an earlier draft.


\providecommand{\bysame}{\leavevmode\hbox to3em{\hrulefill}\thinspace}

\end{document}